\documentclass[12pt]{article}
\usepackage{amsmath}
\usepackage{amssymb}
\usepackage{array}
\usepackage{geometry}
\usepackage{graphicx}
\usepackage{enumerate}
\numberwithin{equation}{section}

\begin{document}
\begin{center}
{\bf \large Analytical results connecting stellar structure parameters and extended reaction rates}\\
\bigskip
{\bf Hans J. Haubold$^{1,2}$ and Dilip Kumar$^2$  }\\
\small{
$^1$Office for Outer Space Affairs, United Nations, \\
Vienna International Centre, P.O. Box 500, A-1400 Vienna, Austria\\
Email: hans.haubold@unoosa.org\\
$^2$Centre for Mathematical Sciences Pala Campus\\
Arunapuram P.O., Palai, Kerala  686 574, India\\
Email: dilipkumar.cms@gmail.com, Website: www.cmsintl.org
}
\end{center}
\begin{center}
{\bf Abstract} \\
\end{center}

{ \small
Possible modification in the velocity distribution in the non-resonant reaction rates leads to an extended reaction rate probability integral.  The closed form representation for these thermonuclear functions are used to obtain the stellar luminosity and neutrino emission rates.  The composite parameter $\mathcal{C}$  that determines the standard nuclear reaction rate through the Maxwell-Boltzmann energy distribution is extended to $\mathcal{C}^*$ by the extended reaction rates through a more general distribution than the Maxwell-Boltzmann distribution.   The new distribution is obtained by the pathway model introduced by Mathai in 2005 [{\it Linear Algebra and Its Applications}, {\bf 396}, 317-328]. Simple analytic models considered by various authors are utilized for evaluating stellar luminosity and neutrino emission rates and are obtained in generalized special functions such as Meijer's $G$-function and Fox's $H$-function. The standard and extended non-resonant thermonuclear functions are compared by plotting them.  Behavior of the new energy distribution, more general than Maxwell-Boltzmann is also studied.\\

}
{\section{\bf Introduction}}
In one way or another, the mystery behind the distant universe is explored by the understanding of the Sun, the star nearest to Earth.  It is the only star whose mass, radius and luminosity are known to highest accuracy.  Structural change in the Sun is due to the evolution of the central gravitationally stabilized solar fusion reactor.  Solar nuclear energy generation and solar neutrino emission are governed by chains of nuclear reactions in this gravitationally stabilized solar fusion reactor \cite{davis2003, hauboldkumar2008}.  Qualitative calculations of specific thermonuclear reaction rates require a large amount of experimental input and theoretical assumption.  By using the theories from nuclear physics and kinetic theory of gases one can determine the reaction rate for low-energy non-resonant thermonuclear reactions in a non-degenerate plasma \cite{hauboldjohn1978}.  The formalization of the calculation of the reaction rate of interacting particles under cosmological or stellar conditions were presented by many authors \cite{fowleretal1967, mathaihaubold1988}. For the most common case, a nuclear reaction in which a particle of type $1$ strikes a particle of type $2$ producing a nucleus $3$ and a new particle $4$ is symbolically represented as
\begin{equation}\label{e12}
1+2 \rightarrow 3+4+ E_{12},
\end{equation}
where $E_{12}$ is the energy release given by $E_{12}= (m_1+m_2-m_3-m_4)c^2$, where $m_i, i=1,2,3,4$ denote the masses of the particles, $c$ denotes the velocity of light. The reaction rate $r_{12}$ of the interacting particles $1$ and $2$ is obtained by averaging the reaction cross section over the normalized density function of the relative velocity of the particles \cite{hauboldmathai1986, mathaihaubold1988, fowler1984}.  Let $n_1$ and $n_2$ denote the number densities of the particles $1$ and $2$ respectively and $\sigma(v)$ be the reaction cross section, where $v$ is the relative velocity of the particles and $f(v)$ is the normalized velocity density, then the reaction rate $r_{12}$ is given by
\begin{align}\label{reactionrate}
r_{12}&=\left(1-\frac{1}{2}\delta_{12}\right)n_1n_2\langle\sigma v\rangle_{12}=\left(1-\frac{1}{2}\delta_{12}\right)n_1n_2 \int_0^\infty v \sigma(v) f(v) {\rm d}v\nonumber \\
&=\left(1-\frac{1}{2}\delta_{12}\right)n_1n_2 \int_0^\infty  \sigma(E) \left(\frac{2E}{\mu}\right)^{\frac{1}{2}} f(E) {\rm d}E,
\end{align}
where $\delta_{12}$ is the Kronecker delta which is introduced to avoid double counting in the reaction, if particles $1$ and $2$ are identical.  The quantity $\langle\sigma v\rangle_{12}$ is the thermally averaged product which is in fact the probability per unit time that two particles $1$ and $2$, confined to a unit volume, will react with each other.  Here, $\mu$ is the reduced mass of the particles given by $\mu=\frac{m_1m_2}{m_1+m_2}$,  $E=\frac{\mu v^2}{2}$ is the kinetic energy of the particles in the center of mass system. From the literature \cite{fowler1984, fowleretal1967, mathaihaubold1988}, it may be noted that all the analytic expressions for astrophysically relevant nuclear reaction rates underline the hypothesis that the distribution of the relative velocities of the reacting particles always remains Maxwell-Boltzmannian for a non-relativistic non-degenerate plasma of nuclei in thermodynamic equilibrium. The Maxwell-Boltzmann relative kinetic energy distribution can be written as
\begin{equation}\label{maxwellboltzmann}
f_{MBD}(E){\rm d}E=2 \pi \left( \frac{1}{\pi kT} \right)^{\frac{3}{2}} \exp\left(-\frac{E}{kT}\right)\sqrt{E}{\rm d} E,
\end{equation}
where $k$ is the Boltzmann constant and $T$ is the temperature.  Substituting (\ref{maxwellboltzmann}) in (\ref{reactionrate}) we get,
\begin{equation}\label{rectionratemaxwellboltzmann}
r_{12}= \left(1-\frac{1}{2}\delta_{12}\right)n_1n_2\left( \frac{8}{\pi \mu} \right)^{\frac{1}{2}}\left( \frac{1}{kT} \right)^{\frac{3}{2}}\int_0^\infty E\sigma(E)\exp\left(-\frac {E}{kT}\right){\rm d}E.
\end{equation}
The thermonuclear fusion depends on three physical variables, the temperature $T$, the Gamow energy $E_G$,  and the nuclear fusion factor $S(E)$. If two nuclei of charges $Z_1e$ and $Z_2e$ collide at low energies below the Coulomb barrier, the Gamow energy $E_G$ is given by \cite{phillips1999,adams2008}
\begin{equation}\label{gamowenergy}
E_G=2\mu(\pi \alpha Z_1Z_2 c)^2,
\end{equation}
where $\alpha$ is the electromagnetic fine structure constant given by
\begin{equation}\label{finestructureconstant}
\alpha=\frac{e^2}{\hbar c},
\end{equation}
where $e$ is the quantum of electric charge, $\hbar$ is Planck's quantum of action, and $\alpha$ is $\frac{1}{137}$\cite{adams2008}.
Thus the Gamow factor, which is determined by the electromagnetic force, and the nuclear fusion factor $S(E)$ determine the nuclear reaction cross section at low energies for non-resonant charged particles as \cite{coradduetal1999, fowler1984}
\begin{equation}\label{crosssection}
\sigma(E)=\frac{S(E)}{E}\exp\left[-\left(\frac{E_G}{ E}\right)^{\frac{1}{2}}\right],
\end{equation}
and $S(E)$ is the cross section factor which is often found to be constant or a slowly varying function of energy over a limited range of energy given by \cite{mathaihaubold1988,fowleretal1967}
\begin{equation}\label{crosssectionfactor}
S(E) \approx  S(0) + \frac{{\rm d}S(0)}{{\rm d}E}E +\frac{1}{2}
 \frac{{\rm d}^2S(0)}{{\rm d}E^2}E^2 =\sum_{\nu=0}^{2}\frac{S^{(\nu)}(0)}{\nu !}E^{\nu}.
 \end{equation}
Substituting (\ref{crosssection}) and (\ref{crosssectionfactor}) in (\ref{rectionratemaxwellboltzmann}) we obtain,
 \begin{align}\label{rijmb}
r_{12}&=\left(1-\frac{1}{2}\delta_{12}\right)n_1n_2 \left( \frac{8}{\pi \mu} \right)^{\frac{1}{2}}\left( \frac{1}{kT} \right)^{\frac{3}{2}}\sum_{\nu=0}^{2}\frac{S^{(\nu)}(0)}{\nu !}\nonumber \\
&\times\int_0^\infty E^\nu\exp\left[-\frac {E}{kT}-\left(\frac{E_G}{ E}\right)^{\frac{1}{2}}\right]{\rm d}E.
\end{align}
This is the non-resonant reaction rate probability integral in the Maxwell-Boltzmann case.  The closed form evaluation of this integral can be seen in a series of papers by Mathai and Haubold, see for example Haubold and Mathai \cite{hauboldmathai1984, hauboldmathai1986}, Mathai and Haubold \cite{mathaihaubold1988}.  The main aim of the present work is to extend the reaction rate probability integral given in (\ref{rijmb}) by replacing the Maxwell-Boltzmann energy distribution by a more general energy distribution called the pathway energy distribution obtained by using the pathway model of Mathai, introduced in 2005.\\

The paper is organized as follows:  In the next section we discuss a more general energy distribution than Maxwell-Boltzmann and obtain the extended reaction rate probability integral in the non-resonant case.  We take advantage of the closed form representation of the extended thermonuclear reaction rate for finding the luminosity and the neutrino emission rate of a so-called non-linear stellar model under consideration in section 3.  Section 4 is devoted to find the desired connection between stellar structure parameters and the neutrino emission of the stellar model by using the closed form analytic representation of the extended reaction rate.  A comparison of the Maxwell-Boltzmann energy distribution with the pathway energy distribution is done with the help of graphs in section 5.  Also we try to discriminate the standard and extended reaction rates.  Concluding remarks are included in section 6.
{\section{\bf Extended non-resonant thermonuclear reaction rate and its closed forms}}

In recent years, possible deviations of the velocity distribution of the plasma particles from the Maxwell-Boltzmann in connection with the production of neutrinos in the gravitationally stabilized solar fusion reactor has been pointed out \cite{coradduetal1999,coradduetal2003, lavagnoquarati2002,lavagnoquarati2006,lissiaquarati2005,hauboldkumar2008}.  It was initiated by Tsallis, the originator of non-extensive statistical mechanics \cite{tsallis1988,tsallis2009,gellmantsallis2004}, who has used the $q$-exponential function as the fundamental distribution instead of the Maxwell-Boltzmann distribution. An initial attempt to extend the standard theories of reaction rates to Tsallis statistics was done by many authors, see Mathai and Haubold \cite{mathaihaubold2007}, Saxena et al. \cite{saxenaetal2004}.  In 2005, Mathai introduced the pathway model by which even more general distributions can be incorporated in the theory of reaction rates \cite{mathai2005, mathaihaubold2007}.  Initially, the pathway model was introduced for the matrix variate case to cover many of the matrix variate statistical densities.  The scalar case is a particular one there. Subsequently, Mathai, his co-workers and others found connection of the pathway model with information theory, fractional calculus, and Mittag-Leffler functions.  The pathway model can be effectively used in any situation in which we need to switch between three different functional forms, namely generalized type-1 beta form, generalized type-2 beta form, and generalized gamma form, using the pathway parameter $q$.  For practical purposes of fitting experimental data, the pathway model can be utilized to switch between different parametric families with thicker or thinner tail.  The pathway model for the real scalar case can be summarized as follows:
\begin{equation}\label{type1beta}
f_1(x)=c_1 x^{\gamma-1}[1-a(1-q)x^\delta]^{\frac{1}{1-q}},~~
a>0,\delta>0, 1-a(1-q)x^\delta>0, \gamma>0,q<1
\end{equation}
is the generalized type-1 beta form of the pathway model.  This is a model with right tail cut-off for $q<1$.  The Tsallis statistics for $q<1$ can be obtained from this model by putting $a=1,\gamma=1,\delta=1$ \cite{gellmantsallis2004,tsallis1988,tsallis2009}.  Other cases available are the regular type-1 beta density, Pareto density, power function, triangular and related models \cite{mathaihaubold2008}.  The generalized type-2 beta-form of the pathway model is given by
\begin{equation}\label{type2beta}
f_2(x)=c_2x^{\gamma-1}[1+a(q-1)x^\delta]^{-\frac{1}{q-1}},~0<x<\infty,q>1,\gamma>0, \delta>0.
\end{equation}
Here, also for $\gamma=1, a=1, \delta=1$,  we get the Tsallis statistics for $q>1$ \cite{gellmantsallis2004,tsallis1988,tsallis2009}. Other standard distributions coming from this model are regular type-2 beta density, $F$-distribution, L\'{e}vy model and related models \cite{mathaihaubold2008}. When  $q \rightarrow 1$,  $f_1 (x)$ and $f_2(x)$ will reduce to the generalized gamma form of the pathway model,,,,, given by
\begin{equation}\label{generalizedgammamodel}
f_3(x)=c_3x^{\gamma-1}{\rm e}^{-a x^\delta},x>0.
\end{equation}
This model covers generalized gamma, gamma, exponential, chisquare, Weibull, Maxwell-Boltzmann, Rayleigh and related densities.  Here, $c_1,c_2$ and $c_3$  are defined in (\ref{type1beta}), (\ref{type2beta}) and (\ref{generalizedgammamodel}), respectively, and are the normalizing constants if we consider statistical densities.

By a suitable modification of the Maxwell-Boltzmann distribution, given in (\ref{maxwellboltzmann}), through the pathway model, we get a more general energy distribution called the pathway energy distribution given by the density
\begin{equation}\label{pathwaydistribution}
  f_{PD}(E){\rm d}E= \frac{2\pi (q-1)^{\frac{3}{2}}}{(\pi kT)^{\frac{3}{2}}} \frac{\Gamma\left(\frac{1}{q-1}\right)}
  {\Gamma\left(\frac{1}{q-1}-\frac{3}{2}\right)} \sqrt{E} \left[1+(q-1)\frac{E}{kT}\right]^{-\frac{1}{q-1}}{\rm d}E,
  \end{equation}
  for $q>1,\frac{1}{q-1}-\frac{3}{2}>0$.  The Maxwell-Boltzmann energy distribution can be retrieved from (\ref{pathwaydistribution}) by taking $q \rightarrow 1$.  Thus the reaction rate probability integral given in (\ref{rectionratemaxwellboltzmann}) can be modified by using (\ref{pathwaydistribution}) and we get the extended reaction rate  as
  \begin{align}
\tilde{r}_{12}&= \left(1-\frac{1}{2}\delta_{12}\right)n_1n_2 \left( \frac{8}{\pi \mu} \right)^{\frac{1}{2}}\left( \frac{q-1}{kT} \right)^{\frac{3}{2}}\frac{\Gamma\left(\frac{1}{q-1}\right)}
  {\Gamma\left(\frac{1}{q-1}-\frac{3}{2}\right)}\sum_{\nu=0}^{2}
  \frac{S^{(\nu)}(0)}{\nu !}\nonumber \\
&\times\int_0^\infty E^\nu \left[1+(q-1)\frac{E}{kT}\right]^{-\frac{1}{q-1}}\exp\left[-\left(\frac{E_G}{ E}\right)^{\frac{1}{2}}\right]{\rm d}E
\end{align}
for $q>1,\frac{1}{q-1}-\frac{3}{2}>0$.  Substituting $y=\frac{E}{kT}$ and $x=\left(\frac{E_G}{kT}\right) ^{\frac{1}{2}}$ we obtain the above integral in a more convenient form
\begin{align}
\tilde{r}_{12}&= \left(1-\frac{1}{2}\delta_{12}\right)n_1n_2 \left( \frac{8}{\pi \mu} \right)^{\frac{1}{2}}(q-1 )^{\frac{3}{2}}\frac{\Gamma\left(\frac{1}{q-1}\right)}
  {\Gamma\left(\frac{1}{q-1}-\frac{3}{2}\right)}\sum_{\nu=0}^{2}
  \left(\frac{1}{kT} \right)^{-\nu+\frac{1}{2}}\nonumber \\
&\times \frac{S^{(\nu)}(0)}{\nu !}\int_0^\infty y^\nu [1+(q-1)y]^{-\frac{1}{q-1}}{\rm e}^{-xy^{-\frac{1}{2}}}{\rm d}y,
\end{align}
for $q>1,\frac{1}{q-1}-\frac{3}{2}>0$.
  Here we consider the integral to be evaluated as
  \begin{equation}\label{i1alpha}
I_{1q}=\int_0^\infty y^\nu [1+(q-1)y]^{-\frac{1}{q-1}}{\rm e}^{-xy^{-\frac{1}{2}}}{\rm d}y.
\end{equation}
The integral can be evaluated by the techniques in applied analysis and can be obtained in closed form via Meijer's $G$-function as \cite{hauboldkumar2008,hauboldkumar2011,kumarhaubold2009}

\begin{equation}\label{i1alphagfunction}
I_{1q}=\frac{(\pi)^{-\frac{1}{2}}}
{(q-1 )^{\nu+1 }\Gamma \left
(\frac{1}{q-1 } \right )}G_{1,3}^{3,1} \left(
\frac{(q-1)x^2}{4} \big|^{2-\frac{1}{q-1}+\nu}_
 {0, \frac{1}{2}, \nu+1} \right)
\end{equation}
which yields the non-resonant reaction rate probability integral in the extended case as \begin{align}\label{r12gnu}
   \tilde{r}_{12}&= \left(1-\frac{1}{2}\delta_{12}\right)n_1n_2 \left( \frac{8}{ \mu} \right)^{\frac{1}{2}}\frac{\pi^{-1}}
  {\Gamma\left(\frac{1}{q-1}-\frac{3}{2}\right)}\sum_{\nu=0}^{2}
  \left(\frac{q-1}{kT} \right)^{-\nu+\frac{1}{2}}\nonumber \\
&\times\frac{S^{(\nu)}(0)}{\nu !}G_{1,3}^{3,1} \left[
\frac{(q-1)E_G}{4kT} \big|^{2-\frac{1}{q-1}+\nu}_
 {0, \frac{1}{2}, \nu+1} \right].
 \end{align}
Meijer's $G$-function and its properties can be seen in Mathai and Saxena \cite{mathaisaxena1973}, Mathai \cite{mathai1993}.  We can obtain series expansions of the $G$-function given in (\ref{r12gnu}) by combining the theories of residue calculus and generalized special functions, see Kumar and Haubold \cite{kumarhaubold2009}, for series expansions for all possible values of $\nu$. In many cases the nuclear factor $S^{(\nu)}(0)$ is approximately constant across the fusion window.  Taking $S^{(\nu)}(0)=0$ for $\nu=1$ and $\nu=2$ and taking $S^0(0)=S(0)$, we obtain the extended reaction rate probability integral as
\begin{equation}\label{r12g}
   \tilde{r}_{12}= \left(1-\frac{1}{2}\delta_{12}\right)n_1n_2 \left[ \frac{8(q-1)}{ \mu kT} \right]^{\frac{1}{2}}\frac{\pi^{-1}}
  {\Gamma\left(\frac{1}{q-1}-\frac{3}{2}\right)}
 S(0)~
  G_{1,3}^{3,1} \left[
\frac{(q-1)E_G}{4kT} \big|^{2-\frac{1}{q-1}}_
 {0, \frac{1}{2}, 1} \right].
 \end{equation}
The series representation for (\ref{r12g}) can be obtained as
\begin{align}\label{r12series}
   \tilde{r}_{12}&= \left(1-\frac{1}{2}\delta_{12}\right)n_1n_2 \left[ \frac{8(q-1)}{ \mu kT} \right]^{\frac{1}{2}}\frac{\pi^{-1}}
  {\Gamma\left(\frac{1}{q-1}-\frac{3}{2}\right)}
 S(0)\bigg\{\sqrt{\pi}\Gamma\left(\frac{1}{q-1}-1\right)\nonumber \\
&-2{\pi}\Gamma\left(\frac{1}{q-1}-\frac{1}{2}\right)
\left[\frac{(q-1)E_G}{4kT}
\right]^{\frac{1}{2}}
{_1F_2}\left(\frac{1}{q-1}-\frac{1}{2};~\frac{3}{2},\frac{1}{2};
~-\frac{(q-1)E_G}{4kT}\right)\nonumber \\
&+\left(\frac{2\sqrt{\pi}(q-1)E_G}{4kT}\right)\sum_{r=0}^{\infty}\left(\frac{(q-1)E_G}{4kT}\right)^r
\bigg[A_r-\ln\left( \frac{(q-1)E_G}{4kT}\right)\bigg] B_r\bigg\}
  \end{align}
where
\begin{equation}
A_r=\Psi\left(-\frac{1}{2}-r\right)+\Psi\left(\frac{1}{q-1}+r \right)+\Psi(1+r)+\Psi(2+r)
\end{equation}
and
\begin{equation}
B_r=\frac{(-1)^{r}\Gamma\left(\frac{1}{q-1}+r\right)
}
{\left(\frac{3}{2} \right)_r r!(1+r)!}.
\end{equation}
see the Appendix for detailed evaluation. As $q\rightarrow1$ in (\ref{r12g}), then by using Stirlings formula for gamma functions, given by
\begin{equation}\label{stirlingsformula}
\Gamma(z+a)\approx(2\pi)^{\frac{1}{2}}z^{z+a-\frac{1}{2}}{\rm e}^{-z},|z| \rightarrow\infty, a \mbox{ is bounded},
\end{equation}
we get the reaction rate probability integral in the Maxwell-Boltzmann case as
\begin{align}
r_{12}&=\left(1-\frac{1}{2}\delta_{12}\right)n_1n_2 \left( \frac{8}{\pi \mu} \right)^{\frac{1}{2}}\left( \frac{1}{kT} \right)^{\frac{3}{2}}S(0)\int_0^\infty \exp\left[-\frac {E}{kT}-\left(\frac{E_G}{ E}\right)^{\frac{1}{2}}\right]{\rm d}E \label{rijmbq1}\\
&=\left(1-\frac{1}{2}\delta_{12}\right)n_1n_2 \left( \frac{8}{\mu k T} \right)^{\frac{1}{2}}S(0)\pi^{-1} G_{0,3}^{3,0} \left[
\frac{E_G}{4kT} \big|^- _
 {0, \frac{1}{2}, 1} \right]\label{rijmbgq1}
\end{align}
which is obtained in a series of papers by Mathai and Haubold, see for example Mathai and Haubold \cite{mathaihaubold1988}.  The integral in (\ref{rijmbq1}) is dominated by the minimum value of $\frac {E}{kT}+\left(\frac{E_G}{ E}\right)^{\frac{1}{2}}=g(E)$ (say).  The minimum value of the function $g(E)$, say $E_0$, can be determined as
\begin{equation}\label{e0}
\frac{{\rm d}}{{\rm d}E}\left[\frac {E}{kT}+\left(\frac{E_G}{ E}\right)^{\frac{1}{2}}\right]_{E=E_0}=\frac{1}{kT}-\frac{1}{2}E_G ^{\frac{1}{2}}E_0 ^{-\frac{3}{2}}=0\Rightarrow E_0=E_G ^{\frac{1}{2}}\left(\frac{kT}{ 2}\right)^{\frac{2}{3}}
\end{equation}
and the function
\begin{equation}\label{ge0}
g(E_0)=3\left(\frac{E_G}{ 4kT}\right)^{\frac{1}{3}}=3\Theta
\end{equation}
where $\Theta=\left(\frac{E_G}{ 4kT}\right)^{\frac{1}{3}}$. Now, by using the Laplace method \cite{Erdelyi1956,olver1974}, we can obtain an approximate value for (\ref{rijmbq1}) as
\begin{equation}\label{approxr12}
r_{12}\approx\left(1-\frac{1}{2}\delta_{12}\right)n_1n_2  \frac{8S(0)\Theta^2\exp(-3\Theta) }{\sqrt{3}\pi \mu \alpha Z_1Z_2 c}.
\end{equation}
  In the next section we will obtain the mass, pressure and temperature  for the case of analytic stellar models characterized by density distribution and corresponding temperature distribution considered by Haubold and Mathai \cite{hauboldmathai1984}.

{\section{\bf Closed forms of the integral over the stellar nuclear energy generation rate}}

Let us consider the density distribution $\varrho(r)$, considered by Haubold and Mathai \cite{hauboldmathai1984,hauboldmathai1986}, Mathai and Haubold \cite{mathaihaubold1988} in the form
\begin{equation}\label{nonlineardensity}
\varrho(r)=\varrho_c \left[1-\left(\frac{r}{R}\right)^\delta \right],\delta>0,
\end{equation}
where $\varrho_c$ is the central density of the star, $r$ is an arbitrary distance from the center and $R$ is the solar radius.  This density function is capable of producing different density distributions by choosing the free parameter $\delta$.  Now we determine the quantities $M(r), P(r)$, and $T(r)$, the mass, pressure and temperature at $r$.

By the equation of the mass conservation,
\begin{equation}
\frac{{\rm d}M(r)}{{\rm d}r}=4 \pi r^2 \varrho(r)
\end{equation}
we get,
\begin{equation}\label{mr}
M(r)=4 \pi \varrho_c \int_0^r t^2 \left[1-\left(\frac{t}{R}\right)^\delta \right]{\rm d} t=\frac{4\pi}{3}\varrho_c r^3 \left[1-\frac{3}{\delta+3}\left(\frac{r}{R}\right)^\delta \right].
\end{equation}
From (\ref{mr}), we get the central density as,
\begin{equation}
\varrho_c=\frac{3 (\delta+3)}{4 \pi \delta}\frac{M(R)}{R^3}.
\end{equation}
If an element of a matter at a distance $r$ from the center of a spherical system is in hydrostatic equilibrium, then setting the sum of the radial forces acting on it to zero we obtain,
\begin{equation}
\frac{{\rm d}P(r)}{{\rm d}r}=-\frac{G \varrho(r)M(r)}{r^2}
\end{equation}
where $G$ is the gravitational constant. Assuming that the pressure at the center of the Sun is $P_c$ and at the surface is zero, we get
\begin{align}\label{prnonlinear}
P(r)&=P_c-G \int_0^r \frac{M(t) \varrho(t)}{t^2} {\rm d} t\nonumber \\
&=\frac{4 \pi}{3} G \varrho_c ^2 R^2 \bigg[ \xi-\frac{1}{2}\left(\frac{r}{R}\right)^2+\frac{\delta+6}{(\delta+2)(\delta+3)}
\left(\frac{r}{R}\right)^{\delta+2}- \frac{3}{2 (\delta+1)(\delta+3)}\left(\frac{r}{R}\right)^{2\delta+2}\bigg].\nonumber \\
\end{align}
 Using the boundary conditions $P(R)=0$, we get,
 $P_c=\frac{4 \pi}{3} G \xi \varrho_c ^2 R^2 $, where
 \begin{equation}\label{xi}
 \xi=\frac{1}{2}-\frac{\delta+6}{(\delta+2)(\delta+3)}
+ \frac{3}{2 (\delta+1)(\delta+3)}.
 \end{equation}
 By the kinetic theory of gases, for a perfect gas, the pressure is given by
 \begin{equation}\label{pressure}
 P(r)=\frac{k N_A}{\mu} \varrho(r)T(r).
 \end{equation}
 For the temperature of interest for stellar models, we neglect the negligible radiation pressure from the total pressure and obtain from (\ref{pressure}) as
 \begin{align}\label{trnonlinear}
 T(r)&=\frac{\mu}{kN_A}\frac{P(r)}{\varrho(r)}\nonumber \\
 &= \frac{4 \pi}{3k N_A} \frac{G \mu\varrho_c  R^2}{[1-(\frac{r}{R})^\delta]} \bigg[ \xi-\frac{1}{2}\left(\frac{r}{R}\right)^2+\frac{\delta+6}{(\delta+2)(\delta+3)}
\left(\frac{r}{R}\right)^{\delta+2}\nonumber \\
&- \frac{3}{2 (\delta+1)(\delta+3)}\left(\frac{r}{R}\right)^{2\delta+2}\bigg].
 \end{align}
 The central temperature is given by $T_c= \frac{4}{kN_A}G \mu \xi \frac{M(R)}{R}$, where $\xi$ is defined in (\ref{xi}).

Thus we have obtained the mass, pressure, and the temperature throughout of the nonlinear stellar model with the density distribution defined in (\ref{nonlineardensity}).  Next our aim is to obtain analytical results for stellar luminosity and neutrino emission rates for various stellar models.
 {\section{\bf Stellar luminosity and neutrino emission rate}}
 The energy conservation equation states that the net increase in the rate of energy flux coming out of a spherical shell from inside is the same as the energy produced with in the shell \cite{hauboldmathai1987}.  If we denote $L_r=L(r)$ as the energy flux through the sphere of radius $r$,  then we have
 \begin{equation}\label{dlr}
 \frac{{\rm d}L_r}{{\rm d}r}=4 \pi r^2 \varrho(r) \varepsilon(r),
 \end{equation}
 where $\varepsilon(r)$ is the energy produced per second by nuclear reactions of each gram of stellar matter.  The quantity $\varepsilon(r)$ depends on the chemical composition in each gram of stellar matter.  Here usually $L_r$ is a constant but will be equal to  $L$ at the surface of the star.  We assume here that the star is chemically homogeneous (that is  a star where chemical composition throughout is a constant).  Also we assume the energy generation rate $\varepsilon(r)$ for one particular nuclear reaction.  Now if we denote $\tilde{r}_{12}(\varrho(r),T(r))$ as the extended non-resonant thermonuclear reaction rate for the particles $1$ and $2$ defined by (\ref{r12g}), then we shall consider the energy generation rate $\varepsilon_{12}(r)$ which can be written in terms of the extended reaction rates via
\begin{equation}\label{epsilon12c}
 \varepsilon_{12}(r)=\frac{1}{\varrho(r)}{\mathcal C} ^{*} \varrho^2 G_{1,3}^{3,1} \left[
\frac{(q-1)E_G}{4kT} \big|^{2-\frac{1}{q-1}}_
 {0, \frac{1}{2}, 1} \right],
\end{equation}
where
\begin{equation}\label{cstar}
 {\mathcal C} ^{*}=\frac{E_{12}\tilde{r}_{12}(\varrho(r),T(r))}{\varrho^2 G_{1,3}^{3,1} \left[
\frac{(q-1)E_G}{4kT} \big|^{2-\frac{1}{q-1}}_
 {0, \frac{1}{2}, 1} \right]},
\end{equation}
in which $E_{12}$ is the amount of energy given off in a single reaction. It is to be noted that by using the asymptotic behavior of $G_{1,3}^{3,1}(\frac{(q-1)E_G}{4kT})$ \cite{mathai1993} and as $q\rightarrow 1$,  ${\mathcal C} ^{*} \rightarrow {\mathcal C}$, the composite parameter considered by \cite{adams2008}, which is defined as
\begin{equation}
 {\mathcal C}=\frac{E_{12}r_{12}}{\varrho^2 \Theta^2 }\exp(3\Theta)
\end{equation}
for our universe $ {\mathcal C}\approx 2\times 10^4$ for proton-proton fusion under typical stellar conditions \cite{adams2008}. Then from (\ref{dlr}) we have the total luminosity of the star by integration as,
 \begin{equation}
 L(R)=\int_0 ^{R} 4 \pi r^2 \varrho(r) \varepsilon(r){\rm d}r.
 \end{equation}
 If we are considering only one specific reaction defined in (\ref{e12}), then we have
 \begin{equation}\label{l12r}
 L_{12}(R)=\int_0 ^{R} 4 \pi r^2 \varrho(r) \varepsilon_{12}(r){\rm d}r,
 \end{equation}
 where the energy generation rate is defined in (\ref{epsilon12c}) and $\varrho(r)$ is a suitable density distribution realized in the sun.  Writing (\ref{l12r}) in terms of $\tilde{r}_{12}(\varrho(r),T(r))$ we get,
 \begin{align}\label{l12rreactionrate}
 L_{12}(R)&=\int_0 ^{R} 4 \pi r^2 {\mathcal C} ^{*}\varrho^2 G_{1,3}^{3,1} \left[
\frac{(q-1)E_G}{4kT} \big|^{2-\frac{1}{q-1}}_
 {0, \frac{1}{2}, 1} \right]{\rm d}r\nonumber\\
&=\int_0 ^{R} 4 \pi r^2 E_{12}\tilde{r}_{12}(\varrho(r),T(r)){\rm d}r.
 \end{align}
 The number density $n_i$ of a particle $i$, for a gas of mean density $\varrho(r)$, can be expressed as
\begin{equation}\label{nir}
n_i(r)=\varrho(r) N_A \frac{X_i}{A_i},
\end{equation}
where $N_A$ stands for Avagadro's constant, $A_i$ the atomic mass of particle $i$ in atomic mass units and $X_i$ is the mass fraction of particle $i$ such that $\sum _i X_i=1$.  Substituting $\tilde{r}_{12}(\varrho(r),T(r))$ from (\ref{r12g}) and using (\ref{nir}) we have
\begin{align}\label{l12rreactionratefinal}
 L_{12}(R)&=\int_0 ^{R} 4 \pi r^2 E_{12}\left(1-\frac{1}{2}\delta_{12}\right)N_A ^2 \varrho ^2(r)  \frac{X_1X_2}{A_1A_2}\left[ \frac{8(q-1)}{ \mu kT(r)} \right]^{\frac{1}{2}}\frac{\pi^{-1}}
  {\Gamma\left(\frac{1}{q-1}-\frac{3}{2}\right)}\nonumber \\
&\times S(0)G_{1,3}^{3,1} \left[
\frac{(q-1)\pi^2\mu}{2kT(r)} \left(\frac{Z_1 Z_2 e^2}{\hbar }\right)^2\big|^{2-\frac{1}{q-1}}_
 {0, \frac{1}{2}, 1} \right]{\rm d}r.
 \end{align}
 If we divide the \textquotedblleft internal luminosity\textquotedblright $L_{12}(R_\odot)$ by the amount of energy $E_{12}$, then we get the total number of particles per second $N_{12}$ liberated in the reaction given by (\ref{e12}) as
 \begin{align}\label{n12final}
 N_{12}&=\frac{L_{12}(R)}{E_{12}}=4  \left(1-\frac{1}{2}\delta_{12}\right)N_A ^2   \frac{X_1X_2}{A_1A_2}\left[ \frac{8(q-1)}{ \mu k} \right]^{\frac{1}{2}}\frac{1}
  {\Gamma\left(\frac{1}{q-1}-\frac{3}{2}\right)}S(0)\nonumber \\
&\times  \int_0 ^{R}\frac{ r^2 \varrho ^2(r)}{[T(r)]^{\frac{1}{2}}}G_{1,3}^{3,1} \left[
\frac{(q-1)\pi^2\mu}{2kT(r)} \left(\frac{Z_1 Z_2 e^2}{\hbar }\right)^2\big|^{2-\frac{1}{q-1}}_
 {0, \frac{1}{2}, 1} \right]{\rm d}r\nonumber \\
 &=4  \left(1-\frac{1}{2}\delta_{12}\right)N_A ^2   \frac{X_1X_2}{A_1A_2}\left[ \frac{8(q-1)}{ \mu k} \right]^{\frac{1}{2}}\frac{1}
  {\Gamma\left(\frac{1}{q-1}-\frac{3}{2}\right)}S(0)\nonumber \\
&\times\frac{1}{2 \pi i}\int_L \Gamma(s) \Gamma\left(\frac{1}{2}+s\right)\Gamma(1+s)\Gamma\left(\frac{1}
{q-1}-1-s\right)
\nonumber \\
&\times \left[\frac{(q-1)\pi^2\mu}{2k}\left( \frac{Z_1Z_2{\rm e}^2}{\hbar}\right)^2\right]^{-s}\int_0 ^{R}r^2
\varrho ^2(r)[T(r)]^{-\frac{1}{2}+s}{\rm d}r {\rm d}s.
\end{align}
 For the density distribution defined in (\ref{nonlineardensity}) introduced by Haubold and Mathai \cite{hauboldmathai1986, mathaihaubold1988} and the corresponding temperature distribution  (\ref{trnonlinear}) we get,
\begin{align}
\int_0 ^{R}r^2 \varrho ^2(r)[T(r)]^{-\frac{1}{2}+s}{\rm d}r
 &=\left[\frac{4 \pi G \mu}{3k N_A}  \varrho_c  R^2 \right]^{s-\frac{1}{2}}\varrho_c ^2\int_0 ^{R} r^2 \left[1-(\frac{r}{R})^\delta\right]^{\frac{5}{2}-s} \bigg[ \xi-\frac{1}{2}\left(\frac{r}{R}\right)^2\nonumber \\
&+\frac{\delta+6}{(\delta+2)(\delta+3)}
\left(\frac{r}{R}\right)^{\delta+2}- \frac{3}{2 (\delta+1)(\delta+3)}
\left(\frac{r}{R}\right)^{2\delta+2}\bigg]^{s-\frac{1}{2}}{\rm d}r,
\end{align}
where $\xi$ is as defined in (\ref{xi}).  If we put a substitution $r=xR$ then we get
\begin{align}
\int_0 ^{R}r^2 \varrho ^2(r)[T(r)]^{-\frac{1}{2}+s}{\rm d}r &=
\left[\frac{4 \pi G \mu}{3k N_A}  \varrho_c  R^2 \right]^{s-\frac{1}{2}}
\varrho_c ^2 R^3 \int_0 ^{1} x^2 [1-x^\delta]^{\frac{5}{2}-s} \bigg[ \xi-\frac{1}{2}x^2\nonumber \\
&+\frac{\delta+6}{(\delta+2)(\delta+3)}
x^{\delta+2}- \frac{3}{2 (\delta+1)(\delta+3)}
x^{2\delta+2}\bigg]^{s-\frac{1}{2}}{\rm d}x.
\end{align}
Putting $x^\delta=y$ and simplifying we obtain,
\begin{align}\label{0torfinal}
\int_0 ^{R}r^2 \varrho ^2(r)[T(r)]^{-\frac{1}{2}+s}{\rm d}r
&=\left[\frac{4 \pi G \mu \xi}{3k N_A}  \varrho_c  R^2 \right]^{s-\frac{1}{2}}\frac{\varrho_c ^2 R^3}{\delta}\nonumber \\
 &\times\int_0 ^{1} y^{\frac{3}{\delta}-1} [1-y]^{\frac{5}{2}-s} [1-u(y)]^{s-\frac{1}{2}}{\rm d}y,
\end{align}
where $u(y)$ is defined as
\begin{equation}
u(y)=\frac{y^{\frac{2}{\delta}}}{\xi}\bigg[ \frac{1}{2}-\frac{\delta+6}{(\delta+2)(\delta+3)}
y+ \frac{3}{2 (\delta+1)(\delta+3)}y^2\bigg].
\end{equation}
As $y \rightarrow 0$, $u(y)\rightarrow 0$ and as $y \rightarrow 1$,
$u(y)\rightarrow 1$.  If we take
\begin{equation}
v(y)=\frac{1}{2}-\frac{\delta+6}{(\delta+2)(\delta+3)}
y+ \frac{3}{2 (\delta+1)(\delta+3)}y^2
\end{equation}
we have $v(0)=\frac{1}{2}$.  The minimum value of $v(y)$ is at $y=\frac{(\delta+1)(\delta+6)}{3(\delta+2)}$ and the value is $\frac{1}{2}-\frac{1}{6}\frac{(\delta+6)^2(\delta+1)}{(\delta+3)(\delta+2)^2}$.  Thus the minimum value is non-negative since $\frac{(\delta+6)^2(\delta+1)}{(\delta+3)(\delta+2)^2}$ decreases steadily from $3$ to $1$ for all $\delta>0$.  Therefore $v(y)\leq0$. Since $\xi>0$ for all $\delta>0$, $u(y)\leq 0$ for all $\delta>0,\xi>0$.
Thus $[1-u(y)]^{s-\frac{1}{2}}\leq 0$.  Hence $0<u(y)<1$ for $0<y<1$ and
 for $\delta>0$.  Thus by using the binomial expansion we obtain
\begin{equation}
[1-u(y)]^{s-\frac{1}{2}}=\sum_{m=0}^\infty \frac{(\frac{1}{2}-s)_m }{m!}[u(y)]^m
\end{equation}
where $(\frac{1}{2}-s)_m $ is the Pochhammer symbol defined for $a \in \mathbb{C}$ by
\begin{align}\label{pochammer}
(a)_ 0=1,(a)_m &=a(a+1)\cdots( a+m-1),m=1,2,\cdots,a\neq 0 \nonumber \\
&=\frac{\Gamma(a+m)}{\Gamma(a)},
\end{align}
 whenever $\Gamma(a)$ exists. Taking $\delta=2$ we get $\xi=\frac{1}{5}$ and
\begin{equation}
[1-u(y)]^{s-\frac{1}{2}}=(1-y)^{2s-1}(1-\frac{1}{2}y)^{s-\frac{1}{2}}.
\end{equation}
Then from (\ref{0torfinal}) we obtain
\begin{align*}
\int_0 ^{R}r^2 \varrho ^2(r)[T(r)]^{-\frac{1}{2}+s}{\rm d}r &=
\left[\frac{4 \pi G \mu }{15k N_A}  \varrho_c  R^2 \right]^{s-\frac{1}{2}}
\frac{\varrho_c ^2 R^3}{2}\nonumber \\
 &\times\int_0 ^{1} y^{\frac{3}{2}-1} [1-y]^{\frac{5}{2}+s-1} (1-\frac{1}{2}y)^{s-\frac{1}{2}}{\rm d}y \nonumber \\
 &=\left[\frac{4 \pi G \mu }{15k N_A}  \varrho_c  R^2 \right]^{s-\frac{1}{2}}\frac{\varrho_c ^2 R^3}{2}\sum_{m=0}^\infty \frac{(\frac{1}{2}-s)_m }{m!} \frac{1}{2^m}\nonumber \\
 &\times\int_0 ^{1} y^{\frac{3}{2}+m-1} [1-y]^{\frac{5}{2}+s-1} {\rm d}y.
 \end{align*}
 By using beta integral and using (\ref{pochammer}) we obtain
 \begin{align}\label{0torfinalanswer}
\int_0 ^{R}r^2 \varrho ^2(r)[T(r)]^{-\frac{1}{2}+s}{\rm d}r &=
\left[\frac{4 \pi G \mu }{15k N_A}  \varrho_c  R^2 \right]^{s-\frac{1}{2}}
\frac{\varrho_c ^2 R^3}{2}\sum_{m=0}^\infty \frac{\Gamma(\frac{1}{2}-s+m) }{m!\Gamma(\frac{1}{2}-s)} \nonumber \\
 &\times \frac{1}{2^m}\frac{\Gamma({\frac{3}{2}+m}) \Gamma({\frac{5}{2}+s})}{\Gamma(4+s+m)} .\end{align}
Now from (\ref{n12final}) we obtain the total number of particles per second
liberated in the reaction (\ref{e12}) as
\begin{align}\label{n12finalsubstitution}
 N_{12}&= \frac{2 N_A ^2 \varrho_c ^2 R^2}{\mu}\left(1-\frac{1}{2}\delta_{12}\right) \frac{X_1X_2}{A_1A_2}\left[\frac{30(q-1)N_A}{\pi G \varrho_c }   \right]^{\frac{1}{2}}S(0)\frac{1}
  {\Gamma\left(\frac{1}{q-1}-\frac{3}{2}\right)} \nonumber \\
  &\times  \sum_{m=0}^\infty \frac{1}{2^m}\frac{\Gamma({\frac{3}{2}+m}) }{m!}
  \frac{1}{2 \pi i}\int_L \Gamma(s) \Gamma(\frac{1}{2}+s)\Gamma(1+s)
  \Gamma({\frac{5}{2}+s})\nonumber \\
&\times\frac{\Gamma\left(\frac{1}{q-1}-1-s\right)
\Gamma(\frac{1}{2}+m-s)
}{\Gamma(\frac{1}{2}-s)\Gamma(4+m+s)}
 \left[\frac{15 \pi(q-1)N_A}{8G\varrho_cR^2}\left( \frac{Z_1Z_2{\rm e}^2}{\hbar}\right)^2\right]^{-s}{\rm d}s
\end{align}
\begin{align}
N_{12}&= \frac{2 N_A ^2 \varrho_c ^2 R^2}{\mu}\left(1-\frac{1}{2}\delta_{12}\right) \frac{X_1X_2}{A_1A_2}\left[\frac{30(q-1)N_A}{\pi G \varrho_c }   \right]^{\frac{1}{2}}S(0)\frac{1}
  {\Gamma\left(\frac{1}{q-1}-\frac{3}{2}\right)} \nonumber \\
  &\times  \sum_{m=0}^\infty \frac{1}{2^m}\frac{\Gamma({\frac{3}{2}+m}) }{m!}  G_{3,5}^{4,2} \left[
\frac{15 \pi(q-1)N_A}{8G\varrho_cR^2}\left( \frac{Z_1Z_2{\rm e}^2}{\hbar}\right)^2\bigg|^{2-\frac{1}{q-1},\frac{1}{2}-m,4+m}_
 {0, \frac{1}{2}, 1,\frac{5}{2},\frac{1}{2}} \right].
\end{align}

For more details on Meijer's $G$-function and its properties see \cite{mathai1993,mathaisaxena1973}.  Thus we have obtained the the total number of particles per second liberated in the reaction (\ref{e12}) in terms of the density distribution considered by Haubold and Mathai \cite{hauboldmathai1986,hauboldmathai1987}.\\
{\section{\bf Comparison of the pathway energy density and the Maxwell-Boltzmann energy density}}

In Figure 1 below it can be observed that for the nuclei to react at energy $E$, it has to borrow an energy $E$ from the thermal environment.  The probability of such an energy is proportional to the Maxwell-Boltzmann energy $\exp\left[-\frac {E}{kT}\right]$.  The fusion will take place when the nuclei penetrate the Coulomb barrier keeping them apart.  The probability of penetration is given by the factor $\exp\left[-\left(\frac{E_G}{ E}\right)^{\frac{1}{2}}\right]$.  The product of these two factors illustrates that fusion mostly occurs in the energy window shown in the following Figure 1.
\begin{center}
\resizebox{12cm}{!}{\includegraphics{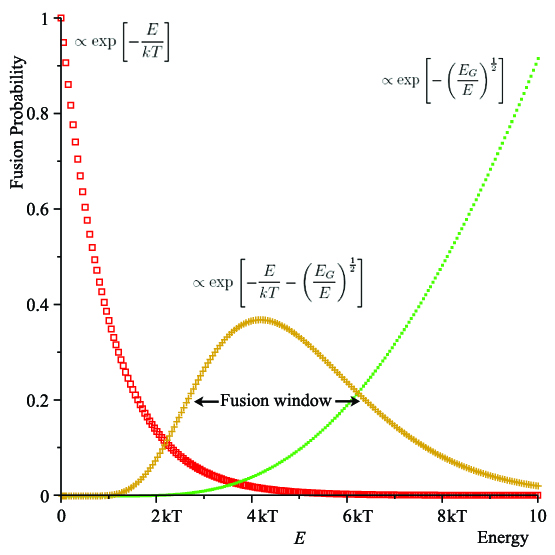}}
\end{center}
\noindent
{\small
{\bf Figure 1:} Schematic plot of the energy-dependent factors for the reaction rate probability integral: Maxwell-Boltzmann energy density, non-resonant nuclear cross-section and the product of the Maxwell-Boltzmann density and the non-resonant cross-section.}\\

In Figure 2 below the pathway energy density is plotted for $q=0.7, 0.9, 1, 1.2, 1.4$, respectively. For different values of $q$ we get different energy densities [Curves {\bf (a)}, {\bf (b)}, {\bf (c)}, {\bf (d)},  {\bf (e)}]. The non-resonant cross-section is also plotted.  The product of the pathway energy density and the non-resonant cross-section for different values of $q$ namely $q=0.7, 0.9, 1, 1.2, 1.4$ are also plotted.  It is to be noted that as $q\rightarrow 1$ the pathway energy density coincides with the Maxwell-Boltzmann energy density and also the fusion window for the Maxwell-Boltzmann case in Figure 1.
\begin{center}
 \resizebox{12cm}{!}{\includegraphics{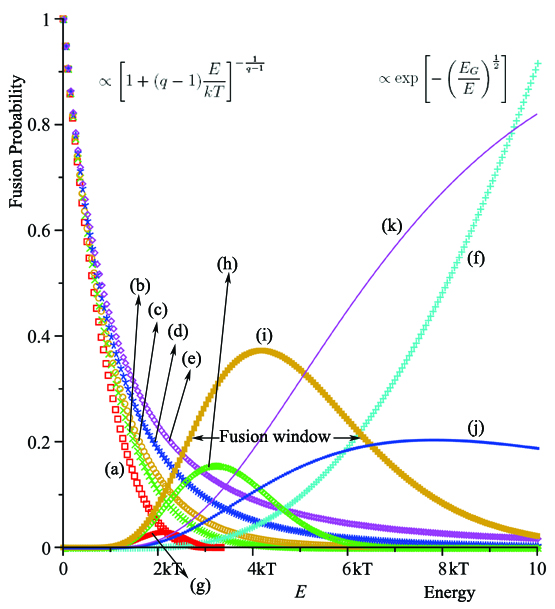}}
\end{center}
\noindent
{\small
{\bf Figure 2:} Schematic plot of the energy-dependent factors for the extended reaction rate probability integral: pathway energy density, non-resonant nuclear cross-section and the product of the pathway density and the non-resonant cross-section.\\
The curves in the figure represents pathway density $\left[1+(q-1)\frac{E}{kT}\right]^{-\frac{1}{q-1}}$ for {\bf (a)} $q=0.7$, {\bf (b)} $q=0.9$, {\bf (c)} $q=1$, {\bf (d)} $q=1.2$, and {\bf (e)} $q=1.4$.
The curve {\bf (f)} represents the non-resonant cross-section $\exp\left[-\left(\frac{E_G}{ E}\right)^{\frac{1}{2}}\right]$. The product $\left[1+(q-1)\frac{E}{kT}\right]^{-\frac{1}{q-1}}\exp\left[-\left(\frac{E_G}{ E}\right)^{\frac{1}{2}}\right]$  is also represented for {\bf (g)} $q=0.7$, {\bf (h)} $q=0.9$, {\bf (i)} $q=1$, {\bf (j)} $q=1.2$, and {\bf (k)} $q=1.4$.\\
}

Figure 3(a) below shows the pathway energy density defined in (\ref{pathwaydistribution}) for $q=1, 1.2, 1.3, 1.4$ and Figure 3(b) shows the Maxwell-Boltzmann energy density defined in (\ref{maxwellboltzmann}).  As $q\rightarrow 1$ the pathway energy density reduces to the Maxwell-Boltzmann energy density.  The pathway energy density covers many stable and unstable situations as the value of $q$ varies.  If the Maxwell-Boltzmann density is the equilibrium situation, many other non-equilibrium situations are covered by the pathway energy density.
\begin{center}
 \resizebox{7cm}{!}{\includegraphics{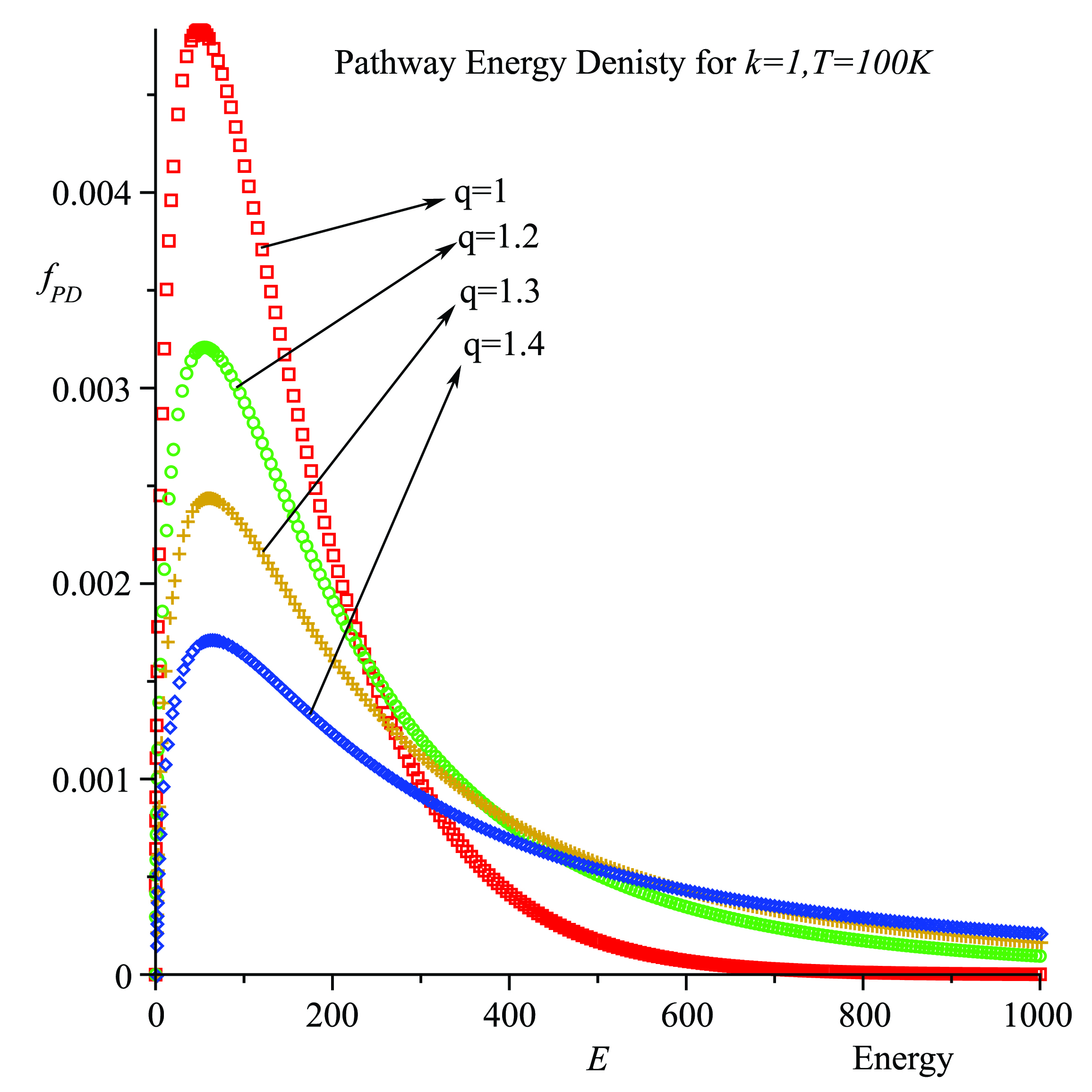}}(a)
 \resizebox{7cm}{!}{\includegraphics{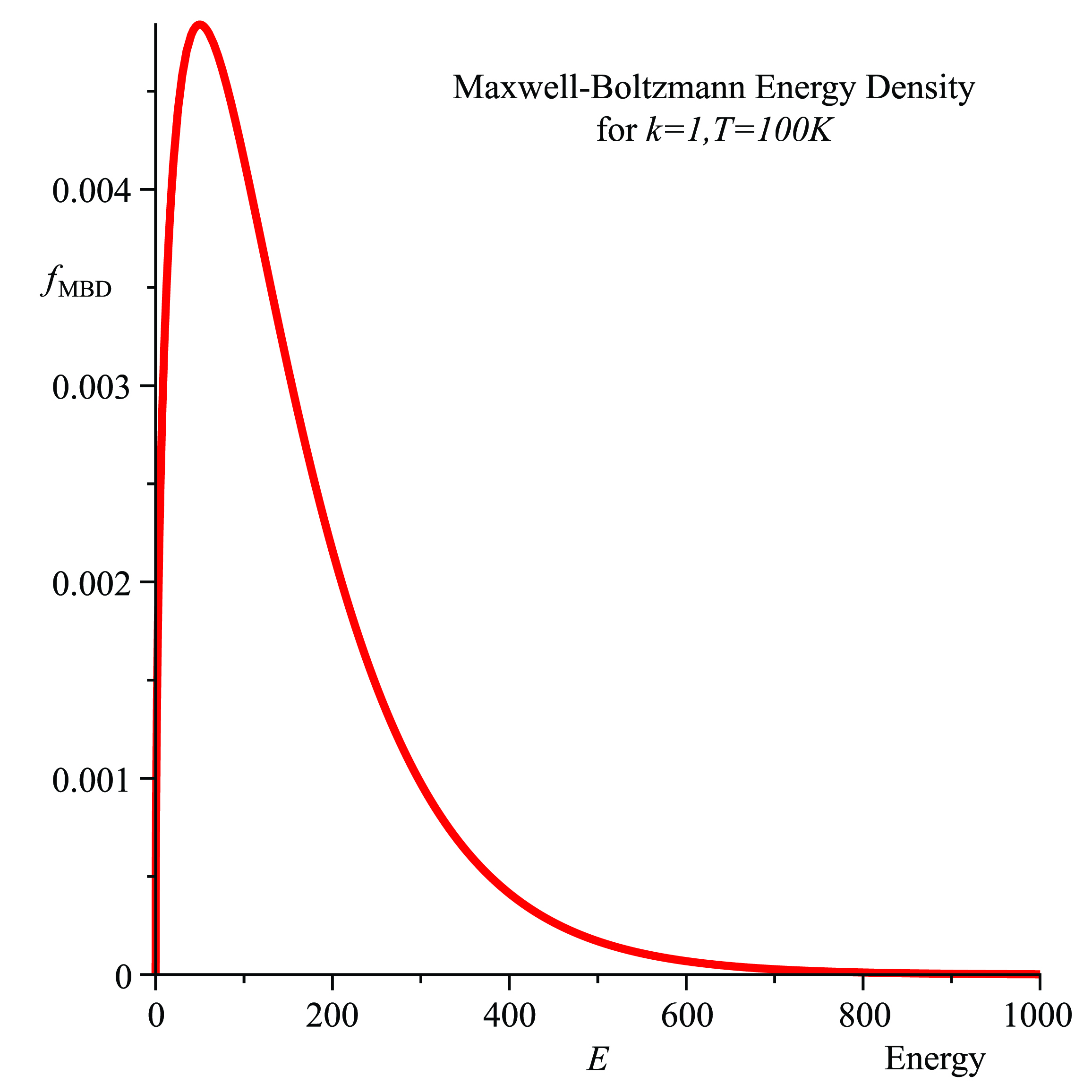}}(b)\\
\noindent
{\small
{\bf Figure 3(a):} Pathway energy density for $q=1, 1.2, 1.3, 1.4$ and for $k=1,T=100K$.\\
{\bf (b):} Maxwell-Boltzmann energy density for $k=1,T=100K$.}
\end{center}
{\section{\bf Concluding remarks}}

In this paper we have modified the energy distribution for a non-resonant reaction rate probability integral.  The composition parameter $\mathcal{C}$ considered by \cite{adams2008} is extended to $\mathcal{C}^*$ by the pathway energy density.  Analytic density distributions, considered by Haubold and Mathai \cite{hauboldmathai1986, hauboldmathai1987}, are used to obtain the stellar luminosity and the neutrino emission rates and are obtained in terms of generalized special functions such as Meijer's $G$-function. The pathway energy density considered here covers many density functions and hence the extended reaction rate integral covers a wider class of integrals.  Pathway energy density helps us to obtain various fusion windows by giving different values to $q$, the pathway parameter.  The graphs plotted here were obtained by using Maple 14 in Windows XP platform.\\

\noindent
{\bf Acknowledgment}\\

The authors would like to thank the Department of Science and Technology,
Government of India, New Delhi, for the financial assistance for this work
 under project No. SR/S4/MS:287/05, and the Centre for Mathematical Sciences
 for providing all facilities.\\
\appendix
{\section{\bf  Appendix}}
\noindent {\bf Series representation}\\

The series representation for the right hand side of (\ref{r12g}) can be obtained through the following procedure.  Here we apply residue calculus on the $G$-function given in (\ref{r12g}).  Consider the $G$-function
\begin{align}
&G_{1,3}^{3,1} \left( \frac{(q-1)E_G}{4kT}
\big|^{2-\frac{1}{q-1}}_
 {0, \frac{1}{2}, 1} \right) = \frac{1}{2 \pi i}
 \int
_{c-i{\infty}}^{c+i{\infty}}\Gamma(s)\Gamma \left(
\frac{1}{2}+s\right)\nonumber \\
&\times  \Gamma (1+s )\Gamma \left ( \frac{1}{q-1 }-1-s \right)
  \left( \frac{(q-1)E_G}{4kT} \right)^{-s} {\rm d}s
 \end{align}
\noindent The right hand side is the sum of the residues of the
integrand. The poles of the gammas in the
integral representation in (\ref{r12g}) are as follows.\\
\medskip
Poles of $\Gamma(s) : ~ s=0,-1,-2,\ldots;$\\
Poles of $\Gamma \left(\frac{1}{2}+s\right) : s =
-\frac{1}{2},-\frac{3}{2},-\frac{5}{2},\ldots ;$\\
Poles of $\Gamma (1+s ):~s=-1,-2,-3,\ldots.$ \\
Here the poles of $\Gamma(s)$ and $\Gamma(1+s)$ will coincide each other at all points except at $s=0$. Note that the pole $s=0$ is a pole of order 1: $s =-\frac{1}{2},-\frac{3}{2},-\frac{5}{2},\ldots$ are each of order 1: $s=-1,-2,-3,\ldots$ are each of order 2.
\noindent
 We know that\\
\begin{eqnarray}
\lim_{s\rightarrow -r}
(s+r)\Gamma(s)&=&\frac{(-1)^r}{r!},\\
\Gamma(a-r)&=&\frac{(-1)^r\Gamma(a)}{(1-a)_r},\\
\Gamma(a+m)&=&\Gamma(a)(a)_m
\end{eqnarray}
when $\Gamma(a)$ is defined,
  $r=0,1,2,\cdots ;~\Gamma
\left(\frac{1}{2}\right)={\pi}^{\frac{1}{2}},$
\\
$(a)_r=\left\{ \begin{array}{ll}
a(a+1)\cdots(a+r-1)& \text{if}~r\geq1,~a\neq0\\
1 & \text{if}~ r=0,\end{array}\right.$\\
The sum of the residues corresponding to the poles
 $s=0$ is given by
\begin{equation}\label{r1final}
R_1=\sqrt{\pi}\Gamma\left(\frac{1}{q-1}-1\right)
\end{equation}
The sum of the residues corresponding to the poles $s =
-\frac{1}{2},-\frac{3}{2},-\frac{5}{2},\ldots $ is
\begin{align}\label{r2final}
R_2&= \sum_{r=0}^\infty \frac{(-1)^r}{r!}\Gamma(-\frac{1}{2}-r)\Gamma(\frac{1}{2}-r)\Gamma(\frac{1}{q-1}-\frac{1}{2}+r)\left[\frac{(q-1)E_G}{4kT}\right]^{-\frac{1}{2}+r}\nonumber \\
&=-2{\pi}\Gamma\left(\frac{1}{q-1}-\frac{1}{2}\right)
\left[\frac{(q-1)E_G}{4kT}
\right]^{\frac{1}{2}}
{_1F_2}\left(\frac{1}{q-1}-\frac{1}{2};~\frac{3}{2},\frac{1}{2};
~-\frac{(q-1)E_G}{4kT}\right)
\end{align}where ${_1F_2}$ is the hypergeometric function defined by
\begin{equation*}
{_1F_2}(a;~b,c;~x)=\sum_{r=0}^{\infty} \frac{(a)_r}{(b)_r (c)_r}
\frac{x^r}{r!}.
\end{equation*}

To obtain the sum of the residues corresponding to poles $~s=-1,-2,-3,\ldots.$ of order 2, we proceed as follows:
\begin{align}\label{r3}
R_3&=\sum_{r=0}^{\infty}\lim_{s\rightarrow-1-r}\frac{\partial}{\partial
s}\bigg[(s+1+r)^2\Gamma(1+s)\Gamma(s)\Gamma\left(\frac{1}{2}+s\right)
\Gamma\left(\frac{1}{q-1}-1-s\right)\left(\frac{(q-1)E_G}{4kT}\right)^{-s}\bigg]\nonumber\\
&=\sum_{r=0}^{\infty}\lim_{s\rightarrow-1-r}\frac{\partial}{\partial
s}\bigg[\frac{\Gamma^2(2+s+r)\Gamma\left(\frac{1}{2}+s\right)\Gamma\left(\frac{1}{q-1}-1-s
\right)}{(s+r)^2(s+r-1)^2
\cdots(s+1)^2 s}\left(\frac{(q-1)E_G}{4kT}\right)^{-s}\bigg]\nonumber\\
&=\sum_{r=0}^{\infty}\lim_{s\rightarrow-1-r}\frac{\partial}{\partial
s}\Phi(s)
\end{align}
where
\[ \Phi(s)=\frac{\Gamma^2(2+s+r)\Gamma\left(\frac{1}{2}+s\right)\Gamma\left(\frac{1}{q-1}-1-s
\right)}{(s+r)^2(s+r-1)^2
\cdots(s+1)^2 s}\left(\frac{(q-1)E_G}{4kT}\right)^{-s}.\]
\noindent
We have
\[\frac{\partial}{\partial s} \Phi(s)=\Phi(s) \frac{\partial}{\partial
s}[ \ln(\Phi(s)] \]
\begin{align*}
\ln\Phi(s)&=2\ln\left[ \Gamma(2+s+r)\right]+\ln\left[
\Gamma\left(\frac{1}{2}+s\right)\right]+\ln\left[
\Gamma\left(\frac{1}{q-1}-1-s \right)\right]\\
&- s \ln \left(\frac{(q-1)E_G}{4kT}\right)-2
\ln(s+r)-2\ln(s+r-1)-\cdots-2\ln(s+1)-\ln(s)\\
\frac{\partial}{\partial s}[
\ln(\Phi(s)]&=2\Psi(2+s+r)+\Psi\left(\frac{1}{2}+s\right)+
\Psi\left(\frac{1}{q-1}-1-s \right)\\
&-\ln\left(\frac{(q-1)E_G}{4kT}\right)-\frac{2}{s+r}-\frac{2}{s+r-1}-\cdots-\frac{2}{s+1}-\frac{1}{s}
\end{align*}
\begin{align}\label{phispartial}
\lim_{s\rightarrow-1-r} \{\frac{\partial}{\partial
s}\ln[\Phi(s)]\} &= \Psi\left(-\frac{1}{2}-r\right)+
\Psi\left(\frac{1}{q-1}+r \right)+\Psi(1+r) \nonumber\\
&+\Psi(2+r)-\ln\left( \frac{(q-1)E_G}{4kT}\right)
\end{align}
where $\Psi(z)$ is a Psi function or digamma function (see Mathai \cite{mathai1993} and $\Psi(1)=-\gamma,
~\gamma =0.5772156649\ldots $ is Euler's constant. Now
\begin{equation}\label{phis}
\lim_{s\rightarrow-1-r}\Phi(s) =\frac{(-1)^{1+r}2\sqrt{\pi}\Gamma\left(\frac{1}{q-1}+r\right)}
{\left(\frac{3}{2} \right)_r r!(1+r)!}\left(\frac{(q-1)E_G}{4kT}\right)^{1 +r}
\end{equation}
Then by using (\ref{r3}), (\ref{phispartial}) and (\ref{phis}) we get,
\begin{align}\label{r3final}
R_3 &=\sum_{r=0}^{\infty}\frac{(-1)^{1+r}2\sqrt{\pi}\Gamma\left(\frac{1}{q-1}+r\right)}
{\left(\frac{3}{2} \right)_r r!(1+r)!}\left(\frac{(q-1)E_G}{4kT}\right)^{1 +r}\nonumber \\
&\times\bigg[\Psi\left(-\frac{1}{2}-r\right)+
\Psi\left(\frac{1}{q-1}+r \right)+\Psi(1+r)+\Psi(2+r)-\ln\left( \frac{(q-1)E_G}{4kT}\right)\bigg]\nonumber\\
&=\left(\frac{2\sqrt{\pi}(q-1)E_G}{4kT}\right)\sum_{r=0}^{\infty}\left(\frac{(q-1)E_G}{4kT}\right)^r
\bigg[A_r-\ln\left( \frac{(q-1)E_G}{4kT}\right)\bigg] B_r,
\end{align}
where
\begin{equation}
A_r=\Psi\left(-\frac{1}{2}-r\right)+\Psi\left(\frac{1}{q-1}+r \right)+\Psi(1+r)+\Psi(2+r)
\end{equation}
and
\begin{equation}
B_r=\frac{(-1)^{r}\Gamma\left(\frac{1}{q-1}+r\right)
}
{\left(\frac{3}{2} \right)_r r!(1+r)!}
\end{equation}
Thus from (\ref{r1final}), (\ref{r2final}) and (\ref{r3final}) we get (\ref{r12series}).
{\small


\begin{thebibliography}{99}
\bibitem{adams2008}
Adams, F.C.: 2008, Stars in other universes: stellar structure with different fundamental constants, {\it Journal of Cosmology and Astroparticle Physics}, {\bf 8}, 1-28.
\bibitem{coradduetal1999}
Coraddu, M., Kaniadakis, G., Lavagno, A., Lissia, M., Mezzorani, G., and Quarati, P.: 1999, Thermal distributions in stellar plasmas, nuclear reactions and solar neutrinos, {\it Brazilian Journal of Physics}, {\bf 29}, 153-168.
\bibitem{coradduetal2003}
Coraddu, M., Lissia, M., Mezzorani, G., and Quarati, P.: 2003, Super-Kamiokande hep neutrino best fit: a possible signal of non-Maxwellian solar plasma, {\it Physica A}, {\bf 326}, 473-481.
\bibitem{davis2003}
Davis Jr., R.: 2003, A half-century with solar neutrinos, {\it Reviews of Modern Physics}, {\bf 75}, 985-994.
\bibitem{Erdelyi1956} Erd\'{e}lyi, A.: 1956, {\it Asymptotic Expansions}, Dover Publications, New York.
\bibitem{fowler1984}
Fowler, W.A.: 1984, Experimental and theoretical nuclear astrophysics: the quest for the origin of the elements, {\it Reviews of Modern Physics}, {\bf 56}, 149-179.
\bibitem{fowleretal1967}
Fowler, W.A., Caughlan, G.R., and Zimmerman, B.A.: 1967, Thermonuclear reaction rates, {\it Annual Review of Astronomy and Astrophysics}, {\bf 5}, 525-570.
\bibitem{gellmantsallis2004}
Gell-Mann, M. and Tsallis, C. (Eds.): 2004, {\it Nonextensive Entropy: Interdisciplinary Applications}, Oxford University Press, New York.
\bibitem{hauboldjohn1978}
Haubold, H.J. and John, R.W.: 1978, On the evaluation of an integral connected with the thermonuclear reaction rate in closed-form, {\it Astronomische Nachrichten}, {\bf 299}, 225-232.
\bibitem{hauboldkumar2008}
 Haubold, H.J. and  Kumar, D.: 2008, Extension of thermonuclear functions through the pathway model including Maxwell-Boltzmann and Tsallis distributions, {\it Astroparticle Physics}, {\bf 29}, 70-76.
  \bibitem{hauboldkumar2011}
 Haubold, H.J. and  Kumar, D.: 2011, Fusion yield: Guderley model and Tsallis statistics, {\it Journal of Plasma Physics}, {\bf 77}, 1-14.
 \bibitem{hauboldmathai1984}
Haubold, H.J. and Mathai, A.M.: 1984, On  nuclear reaction rate theory, {\it Annalen der Physik (Leipzig)}, {\bf 41}, 380-396.
  \bibitem{hauboldmathai1986}
Haubold, H.J. and Mathai, A.M.: 1986, Analytic representations of modified non-resonant thermonuclear reaction rates, {\it Journal of Applied Mathematics and Physics (ZAMP)} {\bf 37}, 685-695.
\bibitem{hauboldmathai1987}
Haubold, H.J. and Mathai, A.M.: 1987, Analytical results connecting stellar structure parameters and neutrino fluxes, {\it Annalen der Physik} {\bf 44}, 103-116.
\bibitem{kumarhaubold2009}
 Kumar, D. and Haubold, H.J.: 2009, On extended thermonuclear functions through pathway model, {\it Advances in Space Research}, {\bf 45}, 698-708.
 \bibitem{lavagnoquarati2002}
Lavagno, A. and Quarati, P.: 2002, Classical and quantum non-extensive statistics effects in nuclear many-body problems, {\it Chaos, Solitons and Fractals}, {\bf 13}, 569-580.
\bibitem{lavagnoquarati2006}
Lavagno, A. and Quarati, P.: 2006, Metastability of electron-nuclear astrophysical plasmas: motivations, signals and conditions, {\it Astrophysics and Space Science}, {\bf 305}, 253-259.
\bibitem{lissiaquarati2005}
Lissia, M. and Quarati, P.: 2005, Nuclear astrophysical plasmas: ion distribution functions and fusion rates, {\it Europhysics News}, {\bf 36}, 211-214.
\bibitem{mathai1993}
Mathai, A.M.: 1993, {\it A Handbook of Generalized Special Functions for Statistics and Physical Sciences}, Clarendon Press, Oxford.
\bibitem{mathai2005}
Mathai, A.M.: 2005, A pathway to matrix-variate gamma and normal densities, {\it Linear Algebra and Its Applications}, {\bf 396}, 317-328.
\bibitem{mathaihaubold1988}
Mathai, A.M. and Haubold, H.J.: 1988, {\it Modern Problems in Nuclear and Neutrino Astrophysics}, Akademie-Verlag, Berlin.
\bibitem{mathaihaubold2007}
Mathai, A.M. and Haubold, H.J.: 2007, Pathway model, superstatistics, Tsallis statistics and a generalized measure of entropy, {\it Physica A}, {\bf 375}, 110-122.
\bibitem{mathaihaubold2008}
Mathai, A.M. and Haubold, H.J.: 2008, On generalized distributions and pathways, {\it Physics Letters A}, {\bf 372}, 2109-2113.
\bibitem{mathaisaxena1973}
Mathai, A.M. and Saxena, R.K.: 1973, {\it Generalized Hypergeometric Functions with Applications in Statistics and Physical Sciences}, Springer, Lecture Notes in Mathematics {\bf Vol. 348}, New York.
\bibitem{olver1974}
Olver, F.W.J.: 1974, {\it Asymptotics and Special Functions}, Academic Press, New York.
\bibitem{phillips1999}
Phillips, A.C.: 1999, {\it The Physics of Stars}, Second Edition, John Wiley \& Sons, Chichester.
\bibitem{saxenaetal2004}
Saxena, R.K., Mathai, A.M., and Haubold, H.J.: 2004, Astrophysical thermonuclear functions for Boltzmann-Gibbs statistics and Tsallis statistics, {\it Physica A}, {\bf 344}, 649-656.
\bibitem{tsallis1988}
Tsallis, C.: 1988, Possible generalization of Boltzmann-Gibbs statistics, {\it Journal of Statistical Physics}, {\bf 52}, 479-487.
\bibitem{tsallis2009}
Tsallis, C. : 2009, {\it Introduction to Nonextensive Statistical Mechanics}, Springer, New York.
\end{thebibliography}
\end{document}